\documentclass[conference]{IEEEtran}
\IEEEoverridecommandlockouts
\usepackage{cite}
\usepackage{amsmath,amssymb,amsfonts}
\usepackage{algorithmic}
\usepackage{graphicx}
\usepackage{textcomp}
\usepackage{xcolor}
\usepackage{booktabs}
\usepackage{graphicx}   
\def\BibTeX{{\rm B\kern-.05em{\sc i\kern-.025em b}\kern-.08em
    T\kern-.1667em\lower.7ex\hbox{E}\kern-.125emX}}
\begin{document}

\title{GUSL-Dehaze: A Green U-Shaped Learning Approach to Image Dehazing}

\author{\IEEEauthorblockN{Mahtab Movaheddrad}
\IEEEauthorblockA{
\textit{University of Southern California}\\
Los Angeles, California, USA \\
movahhed@usc.edu}
\and
\IEEEauthorblockN{Laurence Palmer}
\IEEEauthorblockA{
\textit{University of Southern California}\\
Los Angeles, California, USA \\
palmerla@usc.edu}
\and
\IEEEauthorblockN{C.-C. Jay Kuo}
\IEEEauthorblockA{\
\textit{University of Southern California}\\
Los Angeles, California, USA \\
jckuo@usc.edu }
}

\maketitle

\begin{abstract}
Image dehazing is a restoration task that aims to recover a clear image from a single hazy input. Traditional approaches rely on statistical priors and the physics-based atmospheric scattering model to reconstruct the haze-free image. While recent state-of-the-art methods are predominantly based on deep learning architectures, these models often involve high computational costs and large parameter sizes, making them unsuitable for resource-constrained devices. In this work, we propose GUSL-Dehaze, a Green U-Shaped Learning approach to image dehazing. Our method integrates a physics-based model with a green learning (GL) framework, offering a lightweight, transparent alternative to conventional deep learning techniques. Unlike neural network-based solutions, GUSL-Dehaze completely avoids deep learning. Instead, we begin with an initial dehazing step using a modified Dark Channel Prior (DCP), which is followed by a green learning pipeline implemented through a U-shaped architecture. This architecture employs unsupervised representation learning for effective feature extraction, together with feature-engineering techniques such as the Relevant Feature Test (RFT)\cite{yang2022supervised} and the Least-Squares Normal Transform (LNT)\cite{wang2023enhancing} to maintain a compact model size. Finally, the dehazed image is obtained via a transparent supervised learning strategy. GUSL-Dehaze significantly reduces parameter count while ensuring mathematical interpretability and achieving performance on par with state-of-the-art deep learning models.
\end{abstract}


\section{Introduction}
Haze, caused by light scattering from small particles, reduces visibility and distorts colors, affecting image quality and degrading the performance of higher-level vision tasks such as image recognition and scene understanding. Consequently, dehazing is vital for recovering occluded details in hazy images.

Traditional dehazing methods rely on a physics‐based atmospheric scattering model to describe how the haze forms in an image \cite{narasimhan2003contrast}. The observed haze is viewed as a mixture of scene radiance and scattered ambient light. Therefore, recovering a clear image requires estimating quantities such as the amount of haze at each pixel, the overall airlight color, or the scene depth. Because this inverse problem is underdetermined, most techniques introduce statistical priors to guide these estimates, the dark channel prior (DCP) being a prime example \cite{he2010single}. Despite their practical success, these approaches rest on simplifications that limit their ability to capture more complex, nonhomogeneous scattering phenomena.

Recently, deep learning architectures have come to dominate predictive tasks \cite{fayyazi2025fair,ranjbar2025beyond,torabi2025large}. Current state-of-the-art methods rely on deep learning architectures, which are black-box methods that require large amounts of computational power\cite{guo2022image,abdollahi2024advanced}. Due to these limitations, they are generally not feasible for deployment on resource-constrained devices. In addition, they demand substantial amounts of clear-hazy pairs to learn the ideal mapping. When provided with synthetic hazy data, this is not an issue but becomes a greater issue in real-world applications. 
A new method, termed Green Learning (GL), has been introduced to overcome several limitations of deep learning. Unlike conventional approaches, GL operates without neural networks. This paper presents GUSL-Dehaze, a novel dehazing framework incorporating a Green U-Shaped dehazing pipeline. Our method uses a physics-based model as its foundation and then employs a U-shaped architecture to predict the dehazed images. First, our method adopts a modified DCP as an initial dehazing step, and then it uses a U-shaped architecture to predict the final dehazed images. This approach consists of three modules:

\begin{itemize}
\item \textbf{Modified DCP:} We improve the traditional dark channel prior by incorporating a learning-based approach. This allows the model to adaptively estimate the haze density and improve the transmission map accuracy in challenging scenes.

\item \textbf{Representation Learning:} This module extracts rich, multiscale feature representations from the hazy input using PixelHop units. The extracted features are further refined to select the most relevant components and generate strong secondary features, enhancing the overall representational capacity.

  \item \textbf{Decision Learning:} 
  Guided by a U-shaped architecture, this module integrates an XGBoost regressor to predict the final dehazed image. The combination of structured learning and boosting enables accurate and interpretable decision making with low computational overhead.
\end{itemize} 
\section{Related Works}
 % We can broadly group the dehazing methods into two categories: traditional, prior-based methods and deep learning based methods. 

\subsection{Dehazing}
Traditional dehazing methods are generally based on the atmospheric scattering model (ASM) \cite{narasimhan2003contrast}, a physics-based model of the formation of hazy images. It assumes that a hazy image $I$ is a convex combination of direct attenuation and airlight. Most prior-based methods make an informed estimate of parameters within the ASM supported by statistical observations in the data, allowing them to invert the ASM formulation and recover the clear image. Methods such as the DCP \cite{he2010single} and the color attenuation prior \cite{zhu2015fast} represent prior-based approaches. However, the ASM approximates the complex interactions between particles and light with some problematic simplifications. For example, it assumes uniform haze, which makes it unsuitable for modeling nonhomogeneous haze.
% \subsection{Deep Learning Methods}

Deep learning methods comprise the current SOTA in the dehazing domain. Within this category, we can partition the approaches into physics-aware and physics-unaware. Physics-aware methods consider some aspects of the ASM to guide feature creation or the overall dehazing process. For example, DehazeNet \cite{cai2016dehazenet} and All-in-One-Dehazing Net \cite{li2017aod} regress the medium transmission maps or rely on a reformulation of the ASM, respectively. Although these models improve over pure prior-based methods, they regress intermediate results and use ASM to recover the scene radiance. Thus, they suffer similar issues to those with prior-based methods. To overcome this limitation, other deep learning methods attempt to regress the clear image directly. These models, including FFA-Net \cite{qin2020ffa}, were the next generation of SOTA dehazing. 

Recently, vision transformer architectures, such as DehazeFormer \cite{song2023vision}, have become the most successful dehazing schemes, with variants comprising most of the best-performing models. Other interesting approaches include physics-aware deep learning models, which use the ASM as a guide for discriminant feature creation. For example, $C^2$PNet\cite{zheng2023curricular} uses the ASM in a dual-branch unit, approximating features associated with the atmospheric light and transmission map for more precise dehazing.
\subsection{Green Learning}

Green Learning (GL) \cite{kuo2023green} represents a paradigm shift away from the resource-heavy deep learning frameworks that dominate AI. By forgoing the iterative gradient-based updates of backpropagation, GL attains substantial savings in computation. Instead, it relies on unsupervised feature extraction techniques, most notably the Saab transform \cite{chen2020pixelhop} and its channel-wise variant \cite{chen2019pixelhop} to distill salient representations without the overhead of end-to-end weight tuning.

To further refine its set of characteristics, GL incorporates discriminative selection mechanisms such as the Discriminant Feature Test (DFT) and the Relevant Feature Test (RFT) \cite{yang2022supervised}. These tests isolate the most informative components, streamline the model's inputs, and boost predictive accuracy. Once optimal features are identified, GL leverages powerful learners such as XGBoost \cite{chen2016xgboost} and Subspace Learning Machines (SLM) \cite{fu2024subspace} to build robust classifiers that adapt gracefully across varied datasets and tasks.

A key advantage of GL lies in its lightweight nature. GL delivers scalable solutions suitable for real-world deployment by removing backpropagation and eschewing monolithic end-to-end training. For example, a Green Learning–based demosaicing approach (GID) has been proposed in \cite{movahhedrad2024green}, demonstrating lower computational demands while maintaining high-quality image reconstruction.

\section{PROPOSED METHOD}

\begin{figure*}[htbp]
% \centerline{\includegraphics{xxx.png}}
\centerline{\includegraphics[width=\textwidth]{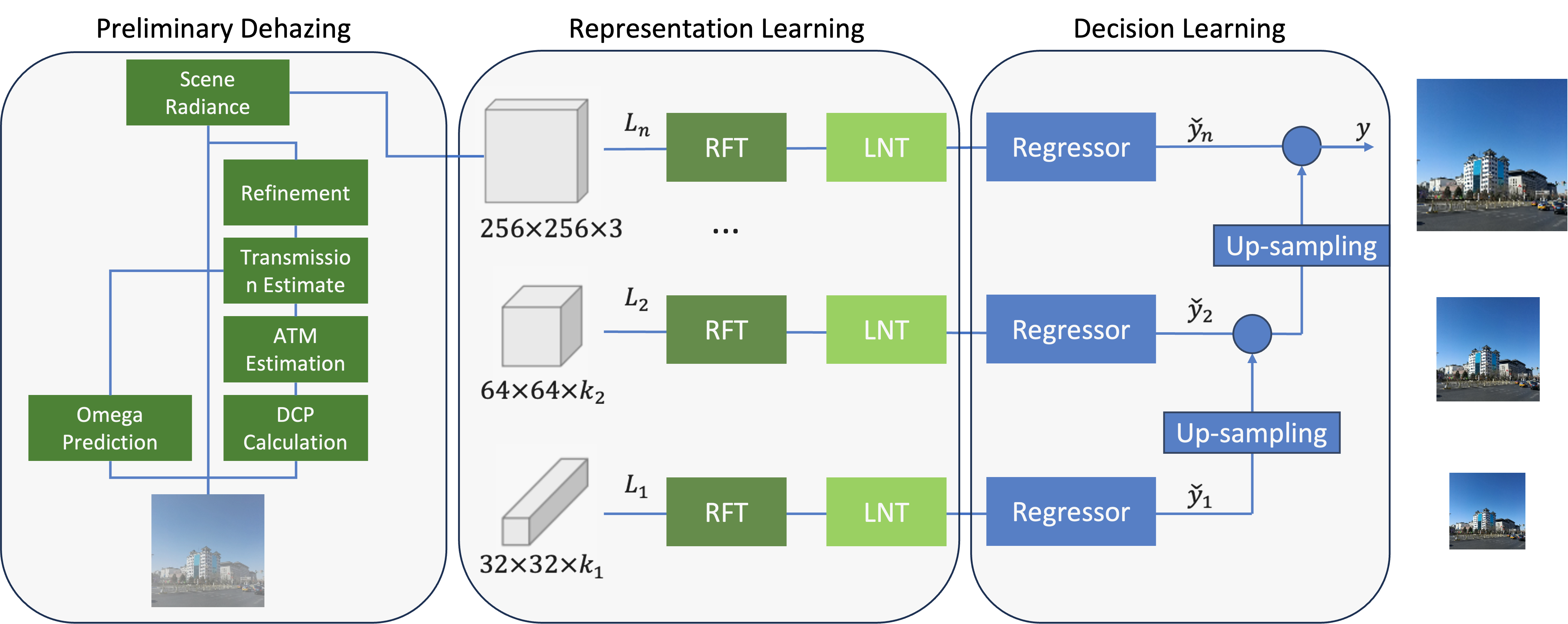}}
\caption{An overview of the proposed pipeline in three modules.
Module 1 represents the preliminary dehazing using a modified DCP method. Module 2 learns robust representations via the Saab transform, selects the most relevant raw features, and then generates secondary features through the LNT.
Module 3 employs two XGBoost regressors at each level to estimate the residuals based on the initial predictions of the preceding (coarser) level.}
\label{fig1}
\end{figure*}

In this section, we present the GUSL-Dehaze method for image dehazing. Our model skips backpropagation altogether. Instead, it iteratively drives down the loss with boosted trees through the XGBoost regressors. The U‑shaped layout in our proposed pipeline calls to mind the U‑Net architecture in many neural networks. However, all parameters within each module are set in a feed‑forward manner. Figure \ref{fig1} illustrates the GUSL-Dehaze block diagram, which is composed of three distinct processing modules. 
Raw images are first projected into a spatial–spectral feature domain during the representation-learning stage. This projection is accomplished through a cascade of PixelHop units that extract and refine latent representations. At each hop, the Saab transform exploits the input patches' intrinsic spatial and spectral attributes, revealing feature representations of increasing abstraction. At each layer, the kernel size is tailored to the image size. Consequently, the filters become progressively smaller in the deeper layers.\\

\subsection{Modified DCP}
The DCP is a prior-based method that relies on the following formulation of the ASM:
\begin{equation*}
    I(x) = J(x)t(x) + A(1-t(x)) 
\end{equation*}
\begin{equation*}
    t(x) = e^{-\beta d(x)}
\end{equation*}
where $x$ is a given pixel, $J$ is the scene radiance, $t$ is the transmission map, $A$ is the airlight, $\beta$ is the atmospheric scattering coefficient, and $d$ is the depth map. The DCP objective is to estimate $t(x)$ and $A$ to recover $J$.

The first step in our pipeline is to perform preliminary dehazing with a modified DCP algorithm. The DCP for a given image $J$ is defined as 
\begin{equation*}
    J^{dark}(x) = \min_{c\in \{r,g,b\}} (\min_{y\in\Omega(x)}(J^c(y)))
\end{equation*}
where $\Omega(x)$ is a patch centered a pixel $x$. Statistically, $J^{dark}$ tends to 0 for clear (outdoor) images. Using this property in the context of the ASM, an estimated transmission map can be recovered using 
\begin{equation*}
    \tilde{t}(x) = 1 - \omega \min_{c \in \{r, g, b\}} \left( min_{y\in\Omega(x)} \left(\frac{I^c(y)}{A^c}\right)\right)
\end{equation*}
The term on the right is the DCP of the normalized haze image $I^c(y)/A^c$ multiplied by $\omega \in (0, 1]$, which is a term to preserve aerial perspective and control the amount of haze removal. $A^c$ is the atmospheric light estimate for the channel $c$. We refine $\tilde{t}(x)$ via a guided filtering algorithm \cite{he2012guided} similar to the soft-matting procedure in the original DCP. Then, the scene radiance can be recovered with 
\begin{equation*}
    J(x) = \frac{I(x) - A}{\max (t(x), t_0)} + A
\end{equation*}
where $t_0$ is a lower bound on the transmission map. The $A$ and the $A^c$ are estimated using the DCP. A percentage of the brightest pixels, $\frac1k$, in the DCP are identified. Among these pixels, we find the pixels with the highest intensity in the input image and use those as our $A$ estimates. 

The DCP has four main parameters of size $\omega$, $t_0$, $\Omega$, and $k$, excluding the guided filter parameters. The $\omega$ has the most significant effect on the recovered image, depending on the intensity of the haze. For example, if $\omega$ is high and the intensity of the haze is low, the recovered image will suffer significant artifacts. Similarly, if $\omega$ is low and the intensity of the haze is high, the recovered image will contain high levels of residual haze. 

To overcome this, we fix all other parameters and regress the $\omega$ value using a random forest model \cite{breiman2001random}. The RESIDE dataset provides $\beta$ parameters associated with the OTS set, which were used to synthesize the hazy images with the ASM equations. These $\ beta$'s serve as a GT of haze intensity. For each $\beta$, we tested for the ideal $\omega$ and trained the random forest model to regress the optimal $\omega$ given the global mean, min, max, and variance within RGB and YUV color spaces. This model generalized well to other datasets while training only on 7K images. 

\subsection{Representation Learning}
\subsubsection{PixelHop feature extraction}PixelHop, introduced in \cite{chen2019pixelhop}, applies the Saab filters to extract features from raw image data. Every PixelHop unit takes a central pixel and its neighborhood \( n\times n\), then flattens them into a vector $\mathbf{v}$.

A Saab filter of dimensions \(m \!\times\! m\) (where \(m < n\))
is applied to this patch. Because the Saab filter is equivalent to a PCA rotation followed by a bias in a successive manner, the patch is projected onto an orthonormal set of principal vectors
\[ \mathbf{U} = \bigl\{\mathbf{u}_{0},\mathbf{u}_{1},\dots, \mathbf{u}_{K-1}\bigr\}, \],
each accompanied by a bias \(c_{k}\).
The \(k\)-th projection coefficient is therefore
\[
z_{k} \;=\;
\mathbf{u}_{k}^{\mathsf{T}}\mathbf{v} \;+\; c_{k},
\qquad k = 0,1,\dots,K-1.
\]

The leading vector \(\mathbf{u}_{0}\) is linked to the largest
eigenvalue and captures the DC (Direct Current) energy of the patch. All remaining vectors \(\mathbf{u}_{1},\dots,\mathbf{u}_{K-1}\) encode the AC (Alternating Current) components.
Hence, the feature space splits orthogonally into
\[
\mathcal{Z}
\;=\;
\mathcal{Z}_{\mathrm{DC}}
\;\oplus\;
\mathcal{Z}_{\mathrm{AC}},
\]
with \(\mathcal{Z}_{\mathrm{DC}} = \operatorname{span}\{\mathbf{u}_{0}\}\)
and
\(\mathcal{Z}_{\mathrm{AC}} = \operatorname{span}\{\mathbf{u}_{1},\dots,
\mathbf{u}_{K-1}\}\).

Let \(\mathbf{v}_{\mathrm{DC}} = \mathbf{u}_{0}^{\mathsf{T}}\mathbf{v} +
c_{0}\) denote the DC projection of the patch.
Subtracting this dominant component yields the residual AC part,
\[
\mathbf{v}_{\mathrm{AC}}
\;=\;
\sum_{k=1}^{K-1}
\Bigl(
\mathbf{u}_{k}^{\mathsf{T}}
\bigl(\mathbf{v}-\mathbf{v}_{\mathrm{DC}}\bigr)
\;+\; c_{k}
\Bigr).
\]

Once the dominant DC energy has been removed, the residual
\(\mathbf{v}_{\mathrm{AC}}\) is subjected to a further PCA so that the remaining variance is concentrated along a compact set of axes.
Let
\(\boldsymbol{\Phi}=\{\boldsymbol{\phi}_{0},\boldsymbol{\phi}_{1},
\dots,\boldsymbol{\phi}_{L-1}\}\) be the orthonormal eigenvectors
obtained from this PCA and let \(d_{l}\) be their associated biases. The refined AC representation is
\[ \tilde{\mathbf{v}}_{\mathrm{AC}}
=\sum_{l=0}^{L-1}
\bigl(\boldsymbol{\phi}_{l}^{\mathsf{T}}\mathbf{v}_{\mathrm{AC}} +d_{l}\bigr)\,\boldsymbol{\phi}_{l}, \]
providing an economical encoding that preserves delicate variations in the data.

At hop \(i\) the algorithm receives an \(n\times n\) neighbourhood.
Convolving this patch with an \(m\times m\) Saab kernel
(\(m<n\)) produces a feature tensor
\[
I^{(i)}\in\mathbb{R}^{N\times N\times K^{\,i-1}},\qquad
N=n-m+1.
\]
Because each hop appends new channels to those of the previous stage, the channel count can grow rapidly. A max‑pool layer is applied immediately after the convolution to maintain tractability, reducing the spatial dimensions while retaining the
strongest responses.
The pooled tensor serves as the input to hop \(i+1\), ensuring that the
hierarchical representation remains both informative and computationally
manageable.
Successive multi‑hop transformations distill a compact and expressive latent embedding that preserves high‑level global context and fine‑grained local structure, providing a stronger foundation for accurate and robust downstream tasks. 

\subsubsection{Relevant Feature Selection}
After the PixelHop feature extraction phase, we have three high-dimensional feature matrices—one per hop. To keep only the most predictive dimensions, we use a supervised feature selection method based on the Relevant Feature Test (RFT) for regression \cite{yang2022supervised}. RFT measures each feature's effect on prediction error and iteratively removes those with little impact, yielding compact, informative representations for each hop.
For each dimension~$j$ and each candidate cut value~$f_t^j$, the RFT partitions the
training data into 
\[
  S_{L,t}^j \;=\;\bigl\{\,\boldsymbol{x}_i \;:\; x_{ij} \le f_t^j \bigr\},
  \quad
  S_{R,t}^j \;=\;\bigl\{\,\boldsymbol{x}_i \;:\; x_{ij} > f_t^j \bigr\}.
\]
Then, we fit separate regression models on the left and right subsets and record their
mean-squared errors~$\mathcal{L}_{L,t}^j$ and~$\mathcal{L}_{R,t}^j$.
The loss associated with threshold~$t$ is the sample-size-weighted average:
\[
  \mathcal{L}_{t}^j \;=\;
  \frac{N_{L,t}^j \,\mathcal{L}_{L,t}^j + 
        N_{R,t}^j \,\mathcal{L}_{R,t}^j}{N},
\]
where $N_{L,t}^j$ and $N_{R,t}^j$ denote the subset sizes and
$N$ is the total number of observations.

The optimal threshold for feature~$j$ minimizes this loss:
\[
  \mathcal{L}_{\min}^j \;=\; 
    \min_{t \in T}\,\mathcal{L}_{t}^j ,
\]
with $T$ the set of all candidate thresholds.
Because smaller values of $\mathcal{L}_{\min}^j$ indicate a cleaner and more predictable split, they serve as a relevance score.

\begin{figure}[htbp]
\centering
\includegraphics[width=\columnwidth]{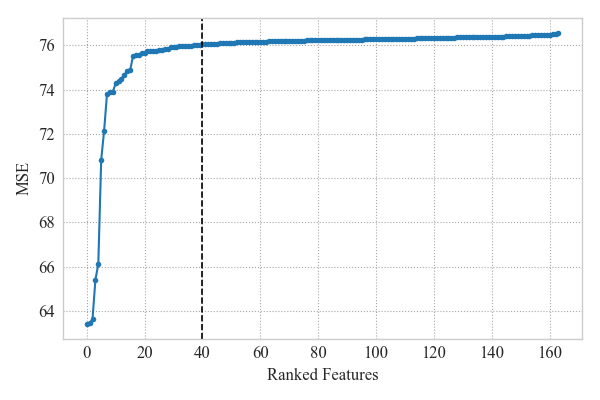}
\caption{Ranked features based on MSE loss using RFT feature selection.}
\label{fig2}
\end{figure}

Finally, we rank all features by $\mathcal{L}_{\min}^j$ in ascending order and retain the $K$ features with the lowest scores. The supervised ranking yields a small subset of variables that improve the performance of the regression.\\
The RFT loss curves reveal which dimensions carry the greatest predictive power. In Figure~\ref{fig2}, the features are sorted by
ascending RFT loss. Thus, variables on the left (low loss) are judged most
informative.
\begin{table}[!t]
  \renewcommand{\arraystretch}{1.3}
  \centering
  \caption{Mean squared error (MSE) for different feature incorporation methods.}
  \label{tab:table1}
  \begin{tabular}{|l|c|c|c|}
    \hline
     & \multicolumn{3}{c|}{MSE} \\ 
    \hline
    Feature incorporated & R     & G     & B     \\ 
    \hline
    Raw                  & 11.35 & 11.04 & 12.72 \\ 
    \hline
    L1 + L2              & 11.44 & 11.28 & 13.03 \\ 
    \hline
    Raw + L1 + L2        & 10.17 &  9.80 & 11.85 \\ 
    \hline
  \end{tabular}
\end{table}
\subsubsection{Feature Generation}
 We selected a subset of the most discriminative features extracted from the images in the previous module. However, expanding the feature space with additional complementary features offers clear advantages. This is particularly important in complex tasks where a limited feature set may not capture all the subtle variations in the data. A diverse feature pool increases the likelihood of capturing relevant patterns.

\begin{table*}[t]
\centering
\caption{Quantitative comparison on various dehazing benchmarks.}
\label{tab:table2}
\resizebox{\textwidth}{!}{%
  \begin{tabular}{@{} 
    l  % Method
    l  % Venue \& Year
    cc % SOTS-indoor: PSNR, SSIM
    cc % SOTS-outdoor: PSNR, SSIM
    cc % Dense-Haze: PSNR, SSIM
    cc % NH-Haze2: PSNR, SSIM
    c  % \#Params
  @{}}
  \toprule
  Method & Venue \& Year 
    & \multicolumn{2}{c}{SOTS-indoor} 
    & \multicolumn{2}{c}{SOTS-outdoor} 
    & \multicolumn{2}{c}{Dense-Haze} 
    & \multicolumn{2}{c}{NH-Haze2} 
    & \#Params \\
  \cmidrule(lr){3-4}\cmidrule(lr){5-6}\cmidrule(lr){7-8}\cmidrule(l){9-10}
   & 
    & PSNR & SSIM 
    & PSNR & SSIM 
    & PSNR & SSIM 
    & PSNR & SSIM 
    &  \\
  \midrule
  DCP \cite{he2010single}         & TPAMI2010    & 16.62 & 0.8179 & 19.13 & 0.8148 & 11.01 & 0.4165 & 11.68 & 0.6475 & –      \\
  AODNet \cite{li2017aod}      & ICCV2017     & 19.06 & 0.8504 & 20.29 & 0.8765 & 12.82 & 0.4683 & 12.33 & 0.6311 & x0.001 \\
    GDN \cite{liu2019griddehazenet}          & ICCV2019     & 32.16 & 0.9836 & 30.86 & 0.9819 & 14.96 & 0.5326 & 19.26 & 0.8046 & x0.56  \\
  GCANet \cite{chen2019gated}       & WACV2019     & 30.06 & 0.9596 & 22.76 & 0.8887 & 12.62 & 0.4208 & 18.79 & 0.7729 & x0.41  \\
  MSBDN \cite{dong2020multi}       & CVPR2020     & 32.77 & 0.9812 & 34.81 & 0.9857 & 15.13 & 0.5551 & 20.11 & 0.8004 & x18.22 \\
  FFA-Net\cite{qin2020ffa}     & AAAI2020     & 36.39 & 0.9886 & 33.57 & 0.9840 & 12.22 & 0.4440 & 20.00 & 0.8225 & x2.59  \\
  AECr-Net\cite{wu2021contrastive}    & CVPR2021     & 37.17 & 0.9901 &  –    &  –     & 15.80 & 0.4660 & 20.68 & \underline{0.8282} & x1.52  \\
DeHamer \cite{guo2022image}     & CVPR2022     & 36.63 & 0.9881 & 35.18 & 0.9860 & \underline{16.62} & 0.5602 & 19.18 & 0.7939 & x77.0 \\
  MAXIM-2S\cite{tu2022maxim}    & CVPR2022     & 38.11 & 0.9908 & 34.19 & 0.9846 &  –    &  –     &  –    &  –     & x8.26  \\

  UDN\cite{hong2022uncertainty}         & AAAI2022     & 38.62 & 0.9909 & 34.92 & 0.9871 &  –    &  –     &  –    &  –     & x2.47  \\
 MB-TaylorFormer-B\cite{qiu2023mb}   & ICCV-2023   & \underline{40.71} & \underline{0.9920} & \textbf{37.42} & 0.9890 &  –    &   –   & \textbf{25.05} & 0.7880 & x1.56 \\ 
  C$^{2}$PNet\cite{zheng2023curricular}      & CVPR-2023           & \textbf{42.56} & \textbf{0.9954} & 36.68 & \textbf{0.9900} & \textbf{16.88} & \textbf{0.5728} & \underline{21.19} & \textbf{0.8334} & x4.17 \\ 
  % DehazeDCT     & CVPR-2024           & 42.56 & 0.9954 & 36.68 & 0.9900 & 16.88 & 0.5728 & 21.19 & 0.8334 & 7.17M \\ 
  
   \midrule
  GUSL-Dehaze      & –            & 39.14 & 0.9911 & \underline{36.83} & \underline{0.9894} & 16.21 & \underline{0.5719} & 20.74 & 0.8154 & x1  \\
  \bottomrule
  \end{tabular}%
}
\end{table*}

We employ the least squares normal transform (LNT)\cite{wang2023enhancing} to obtain these complementary features. This method reformulates the original multi-class classification task into a linear regression framework, deriving discriminative transformations that enhance the features selected in earlier stages.
% --- start of snippet ---
Suppose we have $\ell$ training instances drawn from $\kappa$ distinct classes.  
Introduce an indicator matrix $\mathcal{T}\in\mathbb{R}^{m\times\ell}$ whose entry $\tau_{m,\ell}$ is $1$ when the
sample $\ell$ belongs to superclass $m$ and $0$ otherwise.  
The task is to learn a weight matrix $\mathbf{A}\in\mathbb{R}^{m\times n}$ that
linearly projects the feature matrix $\mathbf{X}\in\mathbb{R}^{n\times\ell}$ onto the
target space, producing the least squares system
\[
\mathbf{A}\mathbf{X} + \mathbf{B} = \mathcal{T},
\]
where $\mathbf{B}$ is a rank one bias term that shifts only the feature mean.
Taking expectations, we can isolate the bias:
\[
\mathbf{B} \;=\; \mathbb{E}[\mathcal{T}] \;-\; \mathbf{A}\,\mathbb{E}[\mathbf{X}].
\]
Because this bias only translates the mean, it does not influence the choice of discriminative directions. The core problem is estimating $\mathbf{A}$.  
Applying the normal equations of linear regression gives
\[
\mathbf{A} \;=\; \mathcal{T}\,\mathbf{X}^{\top}\bigl(\mathbf{X}\mathbf{X}^{\top}\bigr)^{-1}.
\]
Once $\mathbf{A}$ is known, any feature vector $\mathbf{x}\in\mathbb{R}^{n}$
can be mapped to its LNT representation $\mathbf{d}\in\mathbb{R}^{m}$ via
\begin{equation}
\mathbf{d} \;=\; \mathbf{A}\mathbf{x} \;=\; (d_{1},\dots,d_{m})^{\top}.
\end{equation}
We group the features into $n$ categories to improve classification accuracy and use XGBoost decision trees to select the most effective subsets.

Following this pipeline, we refer to the features taken directly from the raw input as Level 1 features. Additionally, we build Level 2 features, not directly from the raw data, but from Level 1 features. This layered approach lets Level 2 features capture more complex patterns and deeper connections, making the overall feature space more expressive and structured.
To evaluate how effective this two-level feature generation is, we compare the MSE results of the validation set by models trained with the raw and combined raw generated features in Table 1.

\subsection{Decision learning}
\subsection*{Decision Learning Module}

The \textbf{Decision Learning} module is the final stage of our dehazing framework, responsible for producing the clean pixel intensity values based on the rich feature representations extracted in earlier stages. This module adopts a hierarchical processing strategy, operating at multiple resolution levels \( \ell \), and utilizes two parallel \textbf{XGBoost regressors} \( f^{(\ell)} \) at each level to handle different feature perspectives.

At each level \( \ell \), the input is a composite feature tensor \( \mathbf{x}^{(\ell)} \in \mathbb{R}^{n \times n \times d} \), where \( n \) denotes the spatial resolution of the current scale, and \( d \) is the number of feature channels. These features include both the \textit{selected raw features} obtained directly from the input and intermediate layers, as well as \textit{synthesized secondary features} derived from feature fusion and transformation processes. These comprehensive representations are designed to encapsulate both low-level texture and high-level semantic information critical for accurate haze removal.

The regression target at this stage is the downsampled ground-truth pixel intensity values, denoted by \( \mathbf{y}^{(\ell)} \in \mathbb{R}^{n \times n \times 1} \), corresponding to a specific color channel. The XGBoost regressors operate independently per channel and are trained to map \( \mathbf{x}^{(\ell)} \) to \( \mathbf{y}^{(\ell)} \), yielding channel-wise dehazed predictions.

XGBoost performs the regression task by minimizing a regularized objective function, defined as:

\[
\mathcal{L}^{(\ell)}(\phi) \;=\; 
\sum_{i=1}^{N} l\!\bigl(y_i^{(\ell)},\,\hat y_i^{(\ell, t)}\bigr)
\;+\;\sum_{k=1}^{t}\Omega\bigl(f_k^{(\ell)}\bigr),
\]

where:
\begin{itemize}
  \item \( l(\cdot,\cdot) \) is a convex loss function (e.g., squared error),
  \item \( \hat y_i^{(\ell, t)} \) is the prediction of the ensemble after \( t \) boosting rounds,
  \item \( f_k^{(\ell)} \) denotes the \( k \)-th regression tree at level \( \ell \),
  \item \( \Omega(f) = \gamma\,T + \frac{1}{2}\,\lambda\|w\|^2 \) is a regularization term that penalizes model complexity, where \( T \) is the number of leaves and \( w \) are the leaf weights.
\end{itemize}

To optimize the objective efficiently, XGBoost employs a second-order Taylor expansion of the loss around the current prediction \( \hat y^{(\ell,t-1)} \):

\[
\mathcal{L}^{(\ell)}\!\approx\!
\sum_{i=1}^{N}\!\Bigl[g_i^{(\ell)}\,f_t^{(\ell)}(x_i^{(\ell)})
+\tfrac12\,h_i^{(\ell)}\,f_t^{(\ell)}(x_i^{(\ell)})^2\Bigr]
+\Omega\bigl(f_t^{(\ell)}\bigr),
\]

where the first and second-order gradients are given by:

\begin{align*}
g_i^{(\ell)} &= \frac{\partial}{\partial\hat y_i^{(\ell,t-1)}}\,l\bigl(y_i^{(\ell)},\hat y_i^{(\ell,t-1)}\bigr), \\
h_i^{(\ell)} &= \frac{\partial^2}{\partial(\hat y_i^{(\ell,t-1)})^2}\,l\bigl(y_i^{(\ell)},\hat y_i^{(\ell,t-1)}\bigr).
\end{align*}

This formulation enables efficient learning with fast convergence and robust generalization. The use of gradient boosting trees allows for both non-linear modeling and interpretability, while the structured multiscale architecture ensures spatial consistency across the image. The modular design also allows the Decision Learning component to remain lightweight, offering low computational overhead without compromising performance.

Overall, this hybrid learning approach in the Decision Learning module enables accurate and context-aware pixel-wise prediction, making it particularly suitable for high-fidelity image dehazing tasks.

\section{Experiments}
We used the RESIDE dataset to train our modified DCP model, which provides synthetic indoor and outdoor images paired with ground truth scattering coefficients ($\beta$ values) for the outdoor training set. The indoor training subset (ITS) comprises 13,990 pairs generated via ground truth depth maps. The Outdoor Training Subset (OTS) contains 72135 pairs synthesized from estimated depth maps, ensuring a diverse range of haze densities and lighting conditions. We initially applied the DCP algorithm to resize hazy images ($256\times 256$) to obtain preliminary dehazed outputs. Based on pre-processed images of size \(256\times256\), we employ a U-shaped network with four successive downsampling stages, generating feature maps at resolutions of \(128\times128\), \(64\times64\), and \(32\times32\). The model predicts residual components at each layer, which are subsequently resampled and integrated into the upper layer to refine the reconstruction. We train two variants of this architecture, one trained in the ITS subset for indoor scenes and another trained in the OTS subset for outdoor environments, and evaluate both in several benchmark datasets. We report the total number of parameters for our proposed model to support our goal of green learning of competitive performance with reduced complexity. Table~\ref{tab:table2} presents our quantitative results alongside parameter counts. Our model achieves PSNR and SSIM scores comparable to state-of-the-art deep learning models, significantly outperforming traditional physics-based dehazing methods. Our model contains 1.72 million parameters in total—approximately 0.43 million per layer, including 0.36 million attributed to the XGBoost regressor. We report the smallest size among the leading deep learning approaches.
 
 \section{Conclusion}

In this work, we proposed a physics-aware green learning framework for image dehazing, GUSL-Dehaze. It delivers performance comparable to state-of-the-art deep learning methods, with the smallest model size reported among them. GUSL-Dehaze is mathematically transparent and computationally efficient. They are well-suited for deployment on edge devices, offering minimal latency, low battery consumption, and efficient memory usage.

% \begin{table}[htbp]
% \caption{Table Type Styles}
% \begin{center}
% \begin{tabular}{|c|c|c|c|}
% \hline
% \textbf{Table}&\multicolumn{3}{|c|}{\textbf{Table Column Head}} \\
% \cline{2-4} 
% \textbf{Head} & \textbf{\textit{Table column subhead}}& \textbf{\textit{Subhead}}& \textbf{\textit{Subhead}} \\
% \hline
% copy& More table copy$^{\mathrm{a}}$& &  \\
% \hline
% \multicolumn{4}{l}{$^{\mathrm{a}}$Sample of a Table footnote.}
% \end{tabular}
% \label{tab1}
% \end{center}
% \end{table}
\section*{Acknowledgment}
The computational work was supported by the University of Southern California Center for Advanced Research Computing (carc.usc.edu).

\bibliographystyle{IEEEtran}
\bibliography{mybib}
\end{document}